\documentclass[useAMS,usenatbib]{mn2e}
\usepackage{amsmath}
\usepackage{amssymb}
\usepackage{graphicx}
\usepackage{subfig}
\usepackage{epsfig}
\usepackage{caption}
\usepackage{breqn}
\usepackage{changebar}
\usepackage{xcolor}

\title[Spectral Distortion by Cumulative CO Emission]{Spectral Distortion of the CMB by the Cumulative CO Emission from Galaxies throughout Cosmic History}
\author[N. Mashian et al.]{Natalie Mashian$^{1}$\thanks{nmashian@physics.harvard.edu}, Abraham Loeb$^{1}$\thanks{aloeb@cfa.harvard.edu}, Amiel Sternberg$^{2}$\thanks{amiel@wise.tau.ac.il} \\
$^{1}$Harvard-Smithsonian Center for Astrophysics, 60 Garden Street, Cambridge, MA 02138, USA\\
$^{2}$The Raymond and Beverly Sackler School of Physics and Astronomy, Tel Aviv University, Tel Aviv 69978, Israel}

\begin{document}

\pagerange{\pageref{firstpage}--\pageref{lastpage}} \pubyear{2015}

\maketitle

\label{firstpage}


\begin{abstract}
We show that the cumulative CO emission from galaxies throughout cosmic history distorts the spectrum of the cosmic microwave background (CMB) at a level that is well above the detection limit of future instruments, such as the Primordial Inflation Explorer (PIXIE). The modeled CO signal has a prominent bump in the frequency interval 100-200 GHz, with a characteristic peak intensity of $\sim$ 2$\times$10$^{-23}$ W\,m$^{-2}$\,Hz$^{-1}$\,sr$^{-1}$. Most of the CO foreground originates from modest redshifts, $z \sim$ 2-5, and needs to be efficiently removed for more subtle distortions from the earlier universe to be detected. 
\end{abstract}



\section{Introduction}
Since the measurements by COBE/Far Infrared Absolute Spectrophotometer (FIRAS), the average cosmic microwave background spectrum (CMB) spectrum is known to be extremely close to a perfect blackbody with a temperature $T_{0}$ = 2.726$\pm$0.001 K and no detected global spectral distortions to date \citep{1994ApJ...420..439M,1996ApJ...473..576F,2009ApJ...707..916F}. 
However, the standard model of cosmology predicts tiny deviations from the Planckian spectrum due to cosmological processes which heat, cool, scatter, and create CMB photons throughout the history of the Universe \citep{1969Natur.223..721S,1969Ap&SS...4..301Z,1970Ap&SS...7...20S,1975SvA....18..413I,1975SvA....18..691I,1982A&A...107...39D,1991A&A...246...49B,1993PhRvD..48..485H,2003MNRAS.342..543B,2012MNRAS.425.1129C,2013IJMPD..2230014S}.
While at redshifts $z\gtrsim$ 2$\times$10$^6$, the thermalization process (mediated by the combined action of double Compton emission, Bremsstrahlung and Compton scattering) is rapid enough to efficiently erase any distortion to unobservable levels, at lower redshifts, the CMB spectrum becomes vulnerable and spectral distortions that form are ``locked in" and can thus be theoretically observed today \citep{2012MNRAS.419.1294C,2014arXiv1405.6938C,2015IJMPD..2430023C}. 

In connection with early energy release, two types of CMB distortions are traditionally distinguished: chemical potential $\mu$- and Compton $y$- distortions \citep{1969Natur.223..721S,1970Ap&SS...7...20S,1975SvA....18..413I,1975SvA....18..691I}. In the regime 2$\times$10$^6 \lesssim z \lesssim$ 3$\times$10$^{5}$, the efficiency of double Compton and Bremsstrahlung processes in controlling the number of CMB photons gradually reduces while photons are still efficiently redistributed in energy by the Compton process. In this case, where thermalization stops being complete, electrons and photons are in kinetic equilibrium with respect to Compton scattering, and any energy injection produces a chemical potential characterized by $\mu(\nu)$. At lower redshifts, $z \lesssim$ 10$^{4}$, up-scattering of photons by electrons also becomes inefficient and photos diffuse only little in energy, creating a $y$-type distortion $y(\nu)$ which is an early-universe analogue of the thermal Sunyaev-Zeldovich effect.
Both types of distortions are tightly constrained by COBE/FIRAS measurements, with upper limits of $|\mu|$ $<$ 9$\times$10$^{-5}$ and $|y|$ $<$ 1.5$\times$10$^{-5}$ at 95\% confidence \citep{1996ApJ...473..576F}.


The amplitude of these signals, predicted within the standard cosmological paradigm, is expected to fall below the bounds set by COBE-FIRAS measurements. The average amplitude of the $y$-parameter, due to the large-scale structure and the reionization epoch, is expected to be $y \simeq$ 10$^{-7}$-10$^{-6}$, with the most recent computations predicting $y \simeq$ 2$\times$10$^{-6}$ \citep{1994PhRvD..49..648H,2000PhRvD..61l3001R,2003MNRAS.342L..20O,2015arXiv150701583H}. 
The $\mu$-distortion signal is expected to be even weaker, with the damping of small-scale acoustic modes giving rise to $\mu \simeq$ 2$\times$10$^{-8}$ in the standard slow-roll inflation scenario \citep{1991ApJ...371...14D,1994ApJ...430L...5H,2012MNRAS.425.1129C}. Although these distortions are small, significant progress in technology in the last two decades promises to detect these spectral distortions. Experimental concepts, like the Primordial Inflation Explorer (PIXIE; \citealt{2011SPIE.8146E..0TK}) and Polarized Radiation Imaging and Spectroscopy Mission (PRISM; \citealt{2014JCAP...02..006A}), could possibly improve the absolute spectral sensitivity limits of COBE/FIRAS by 2-3 orders of magnitude and detect the aforementioned signals at the 5$\sigma$ level, providing measurements at sensitivities $\mu$ = 5$\times$10$^{-8}$ for a chemical potential distortion and $y$ = 10$^{-8}$ for a Compton distortion \citep{2002ApJ...581..817F, 2011SPIE.8146E..0TK,2013MNRAS.436.2232C,2014MNRAS.438.2065C}.

However, it is not yet clear what the foreground limitations to measuring these primordial spectral distortions will be. When considering the large angular scales of interest to PIXIE, focus has been geared towards the foreground subtraction of polarized emission from the Milky Way's interstellar medium (ISM) which is dominated by synchrotron radiation from cosmic ray electrons accelerated in the Galactic magnetic field, and thermal emission from dust grains. \citet{2011JCAP...07..025K} claim that the CMB emission can be distinguished from Galactic foregrounds based on their different frequency spectra, as long as the number of independent frequency channels equals or exceeds the number of free parameters to be derived from a multi-frequency fit. 

But in addition to these Galactic foregrounds, there is another contaminant which has been primarily neglected in the literature, and that is the diffuse background of CO emission lines from external galaxies.  Until recently, \citet{2008A&A...489..489R} provided the only estimate of this redshift-integrated CO emission signal. Assuming star formation is driven by major mergers, they calculated the resulting star-formation rate (SFR) and then converted it to a CO luminosity, $L_{CO}$, using the measured ratio of $L_{CO}$ to SFR in M82, a low-redshift starburst galaxy. The CO background they found, integrated over all redshifts, is expected to contribute $\sim$ 1 $\mu$K at $\nu \gtrsim$ 100 GHz with an almost flat spectrum.
\citet{2015arXiv151204816D} also include estimates of the background contributed by CO lines from star-forming galaxies when they consider the Galactic and extragalactic foreground intensity compared with the CMB spectra. Using the $L_{\text{IR}}$-$L_{\text{CO}}$ relations presented in \citet{2014ApJ...794..142G} for the CO rotational ladder from $J$ = 1$\rightarrow$ 0 to $J$ = 5$\rightarrow$ 4, they find that the CO signal is substantially higher than the PIXIE sensitivity, with the CO(4-3) line alone contributing $\sim$ 3$\times$10$^{-24}$ erg s$^{-1}$ cm$^{-2}$ sr$^{-1}$ at sub-mm wavelengths.

In this paper, we apply the formalism and machinery presented in \citet{2015JCAP...11..028M} to predict the total CO emission signal generated by a population of star-forming halos with masses $M\geq$ 10$^{10}$ M$_\odot$ from the present-day, to redshifts as early as $z \sim$ 15. 
Our comprehensive approach is based on large-velocity gradient (LVG) modeling, a radiative transfer modeling technique that produces the full CO spectral line energy distribution (SLED) for a specified set of parameters characterizing the emitting source. By linking these parameters to the global properties of the host halos, we calculate the CO line intensities emitted by a halo of mass $M$ at redshift $z$, and then further integrate these CO luminosities over the range of halo masses hosting CO-luminous galaxies to derive the average surface brightness of each rotational line. We find that over a range of frequencies (30-300 GHz) spanned by a PIXIE-like mission, the signal strength of this diffuse background is 1-3 orders of magnitude larger than the spectral distortion limits PIXIE aims to provide. The CO foreground must be removed in order for the more subtle distortion signals from the earlier universe to be detected.

This Letter is organized as follows. Section 2 provides a brief overview of the formalism and key ingredients of our CO-signal modeling technique. Section 3 presents the results and Section 4 summarizes the main conclusions. We adopt a flat, $\Lambda$CDM cosmology with $\Omega_m$ = 0.3, $\Omega_\Lambda$ = 0.7, $\Omega_b$ = 0.045, $H_0$ = 70 km s$^{-1}$Mpc$^{-1}$ (i.e. $h$ = 0.7), $\sigma_8$ = 0.82, and $n_s$ = 0.95, consistent with the most recent measurements from Planck \citep{2015arXiv150201589P}.

\section{The Formalism}

In \citet{2015JCAP...11..028M}, we developed a novel approach to estimating the line intensity of any CO rotational transition emitted by a host halo with mass $M$ at redshift $z$ in the early universe, $z \geq$ 4. Here, we briefly outline the model far enough to calculate the quantities relevant for the present work and refer the reader to our previous paper for further details.

In our formalism, the average specific intensity of a given CO line with rest-frame frequency $\nu_J$ emitted by gas at redshift $z_J$ is,
\begin{equation}
I_{\nu_{obs}}=\frac{c}{4\pi}\frac{1}{\nu_J H(z_J)}\int_{M_{min,CO}}^\infty\hspace{-0.7cm} dM\,\,\frac{dn}{dM}(M,z_J)L(M,z_J)
\end{equation}
where $H(z)$ is the Hubble parameter, $dn/dM$ is the Sheth-Tormen halo mass function \citep{1999MNRAS.308..119S} and $M_{min,CO}$ is the minimum host halo mass for CO-luminous galaxies. To determine the specific luminosity of the line, $L(M,z_J)$, we employ LVG modeling, a method of radiative transfer in which the excitation and opacity of CO lines are determined by the kinetic temperature $T_{\rm{kin}}$, velocity gradient $dv/dr$, gas density $n$, CO-to-H$_2$ abundance ratio $\chi_{\rm{CO}}$, and the CO column density $N_{\text{CO}}$ of the emitting source. 
A background radiation term with temperature, $T_{CMB}=T_0(1+z)$, is included in the LVG calculations; the increasing CMB temperature with $z$ is expected to depress the CO line luminosity at higher redshifts.
 Adopting the escape probability formalism \citep{1970MNRAS.149..111C,1974ApJ...189..441G} for a spherical cloud undergoing uniform collapse and assuming that each emitting source consists of a large number of these unresolved collapsing clouds, the emergent LVG-modeled intensity of an emission line can be expressed as
\begin{equation}
I_J=\frac{h\nu_J}{4\pi} A_J x_J\beta_J(\tau_J)\chi_{CO}N_{H_2}
\end{equation}
where $x_J$ is the population fraction in the $J^{th}$ level, $A_J$ is the Einstein radiative coefficient, $N_{H_2}$ is the beam-averaged H$_2$ column density, and $\beta_J = (1-e^{\tau_{J}})/\tau_{J}$ is the photon-escape probability. To carry out these computations, we use the Mark \& Sternberg LVG radiative transfer code described in \citet{dave}.

 We showed previously that the LVG parameters, \{$T_{\rm{kin}}$, $n_{\rm{H_2}}$, $dv/dr$, $\chi_{\rm{CO}}$, and $N_{\rm{H_2}}$\} , which drive the physics of CO transitions and ultimately dictate both the shape and amplitude of the resulting CO SLED, can be linked to the emitting galaxy's global star formation rate, $SFR$, and the star formation rate surface density, $\Sigma_{SFR}$. The analytic expressions for these quantities can be found in Section 2.3 of \citet{2015JCAP...11..028M}, and will not be rederived here. 
The final ingredient in our model is thus a $SFR-M$ relation that allows us to express these LVG parameters solely as functions of the global properties of the host halo, i.e. the halo mass $M$ and redshift $z$.

For high redshifts, $z \geq$ 4, we adopt the average $SFR-M$ relation derived in \citet{2016MNRAS.455.2101M} via abundance-matching. Assuming each dark-matter halo hosts a single galaxy, they mapped the shape of the observed ultraviolet luminosity functions (UV LFs) at $z \sim$ 4-8 to that of the halo mass function at the respective redshifts and found that the $SFR-M$ scaling law is roughly constant across this redshift range (within 0.2 dex). This average relation, which faithfully reproduces the observed $z \sim$ 9 - 10 LFs, is therefore employed in our calculations for all redshifts greater than 4. 
For $z <$ 4, we rely on the results of \citet{2013ApJ...770...57B,2013ApJ...762L..31B} which empirically quantified the stellar mass history of dark matter halos, using a comprehensive compilation of observational data along with simulated halo merger trees to constrain a parameterized stellar mass-halo mass relation. 

\begin{figure}
\hspace{-1.2cm}\includegraphics[width=105mm,height=80mm]{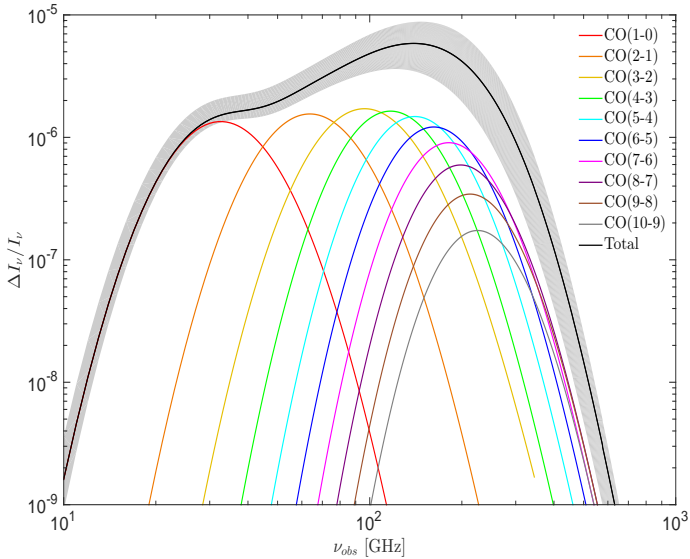}
\caption{The relative CMB spectral distortions due to CO emission from star-forming galaxies at redshifts as high as $z \sim$ 15 to the present. The contribution of each spectral line, from J = 1$\rightarrow$ 0 to J = 10$\rightarrow$ 9 is shown, along with the summed signal (black curve). The shaded area corresponds to the 1$\sigma$ confidence region for the total predicted signal.  }
\end{figure}

\begin{figure}
\hspace{-.8cm}\includegraphics[width=105mm,height=80mm]{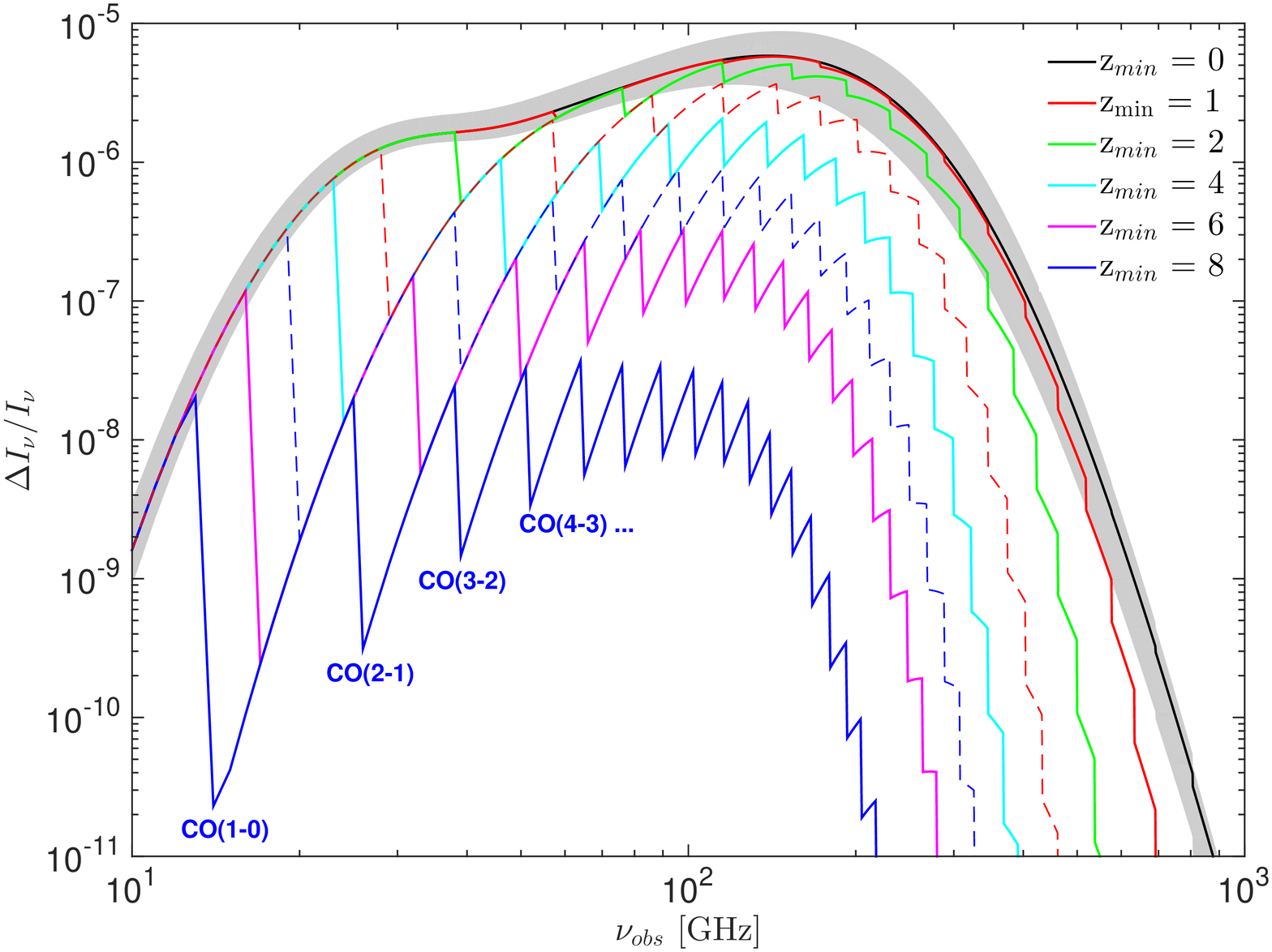}
\caption{The relative CMB spectral distortions due to the cumulative CO foreground emission from galaxies at redshifts between $z = z_{\rm{min}}$ and $z$ = 15. }
\end{figure}

\section{Results}
In Figure 1, we present our predictions of the contribution of each spectral line to the cumulative CO background from star-forming galaxies at redshifts as high as $z \sim$ 15 to the present.
The low-$J$ CO lines peak at frequencies corresponding to an emission redshift of $z \sim$ 2. This emission is dominated by star-forming halos with masses 10$^{11}$-10$^{12}$ M$_\odot$, hosting molecular clouds that are characterized by a gas kinetic temperature $T_{\rm{kin}} \sim$ 40 K and H$_2$ number densities $n_{\rm{H_2}} \sim$ 100 - 1000 cm$^{-3}$.
The higher-$J$ ($J >$ 5) CO signal is dominated by emission from 10$^{11}$-10$^{12}$ M$_\odot$ host halos residing at $z \gtrsim$ 4; in the star-forming galaxies at these high redshifts, the physical conditions in the emitting molecular clouds are extreme enough to thermalize the higher-order CO transitions, with gas kinetic temperatures of $\sim$ 60 K and number densities of order 10$^4$ cm$^{-3}$. 
Integrating over the population of CO luminous halos between redshifts 0 $\leq z\leq$ 15, we find that the total emission (black curve) predicted by our LVG-based model is not completely spectrally smooth, but rather has a prominent bump in the frequency interval $\sim$ 100 - 200 GHz with a characteristic peak intensity of $\sim$ 2$\times$10$^{-23}$ W\,m$^{-2}$\,Hz$^{-1}$\,sr$^{-1}$, i.e. $\Delta I_\nu/I_\nu \simeq$ (5$\pm$2)$\times$10$^{-6}$. 
This is the frequency range within which the most prominent redshifted CO line emissions, originating from sources at redshifts $z \sim$ 2 - 5, fall and accumulate to form the peak in the cumulative signal depicted in Figure 1.
The total CO intensity is $\sim$ 0.01\% of the far-infrared background intensity computed in \citet{1998ApJ...508..123F} and \citet{2014ARA&A..52..373L}, where the former computes a total 125-2000 $\mu$m background of $\sim$ 14 nW\,m$^{-2}$sr$^{-1}$ and the latter computes a total 8-1000 $\mu$m background of $\sim$ 27 nW\,m$^{-2}$sr$^{-1}$.
The uncertainty in our estimates of the CO signal, represented by the shaded regions in Figures 1-3, primarily stems from the uncertainty in the average $SFR-M$ relations we adopt to express the LVG parameters as functions of the global properties of the host halos.

In the case where we assume that local sources of CO emission can be identified and subtracted from observations, we find that the predicted foreground signal not only drops in magnitude as expected, but the shape of the CO spectrum is modified as well. These results are clearly demonstrated in Figure 2, where each curve is computed by integrating the CO intensity emitted by galaxies from some minimum redshift, $z_{\rm{min}}$, out to redshift $z \simeq$ 15. Starting off with $z_{\rm{min}}$ = 0, which corresponds to the original results shown in Figure 1 where emission from local sources is included in the calculations (black curve), we vary the minimum redshift to values as high as $z_{\rm{min}}$ = 8.  We find that excluding the emission from the population of lower redshift sources results in a sawtooth modulation of the cumulative CO spectrum, with the modulation appearing at the observed frequencies, $\nu_{\rm{obs}} = \nu_J/(1+z_{\rm{min}})$, at which the contribution from a given CO transition, $J \rightarrow J-1$, drops out. For example, in the case where emission sources at redshifts $z <$ 4 can be subtracted and thus no longer contribute to the CO foreground (blue curve), the total CO signal strength plummets at $\nu_{\rm{obs}} \simeq$ 23 GHz when the CO(1-0) line contribution disappears, and then again at $\nu_{\rm{obs}} \simeq$ 46 GHz when the CO(2-1) contribution dies out; this pattern continues out to $\nu_{\rm{obs}} \simeq$ 922 GHz, at which point the redshifted CO($J$ = 40 $\rightarrow$ 39) line vanishes and no trace of the CO signal emitted by the galaxy population at $z \geq$ 4 is left. This sawtooth-shaped form of the resulting CO spectrum highlights both the discrete nature of CO rotational transitions at frequencies, $\nu_J = J\nu_{\rm{CO(1-0)}}$, with $\nu_{\rm{CO(1-0)}}$ = 115.3 GHz and $J$ = 1, 2, ..., 40, as well as the significant contribution by modest redshift sources to the overall CO foreground signal.

\begin{figure}
\hspace{-.8cm}\includegraphics[width=105mm,height=80mm]{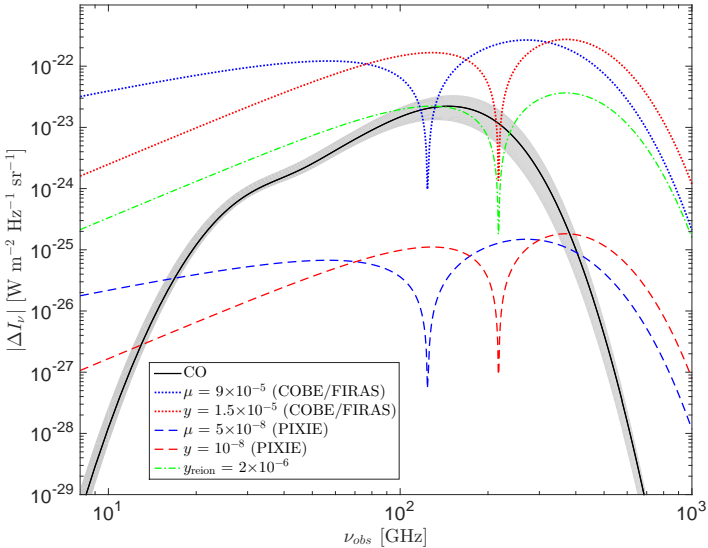}
\caption{$\mu$-type and $y$-type spectral distortions corresponding to the current COBE/FIRAS limits (2$\sigma$; dotted curves) and the anticipated PIXIE sensitivity limits (5$\sigma$; dashed curves). The absolute value of the difference in intensity from the blackbody spectrum is shown where $\Delta I_\nu^\mu$ and $\Delta I_\nu^y$ are given by equations (3) and (4), respectively. The green dash-dot curve represents the $y$-type distortions due to reionization and structure formation in the late Universe, $z \lesssim$ 10 - 20. The cusp in each curve signifies the transition from a negative distortion to a positive distortion, with a zero point/crossing frequency of $\nu$ = 124 and 217 GHz for $\mu$- and $y$-type distortions, respectively.}
\end{figure}

Assuming that the more local emission sources are not individually subtracted, the intensity of the predicted CO background is at least 2-100 times weaker than the current COBE/FIRAS upper limits, $\Delta I_\nu/I_\nu \lesssim$ 10$^{-5}$ - 10$^{-4}$. Therefore, although this foreground is expected to peak in the range of frequencies spanned by COBE/FIRAS, the signal has eluded detection to date. However, as depicted in Figure 3, the total emission one expects from the CO background lies 1-3 orders of magnitude above the PIXIE sensitivity to $\mu$- and $y$-type distortions in the 30-300 GHz frequency range. 
These spectral distortions take the form
\begin{equation}
\Delta I_\nu^\mu = \frac{2h\nu^3}{c^2}\times \mu\frac{e^x}{(e^x-1)^2}\left(\frac{x}{2.19}-1\right)
\end{equation}
and
\begin{equation}
\Delta I_\nu^y = \frac{2h\nu^3}{c^2}\times y\frac{xe^x}{(e^x-1)^2}\left[x\left(\frac{e^x+1}{e^x-1}\right)-4\right]\end{equation}
where $x = h\nu/(k_BT)$ is the dimensionless frequency, $h$ is Planck's constant, $k_B$ is Boltzmann's constant, and $T$ is the CMB temperature. 
The full exploitation of PIXIE's potential to measure $\mu$- and $y$-type spectral distortions of $\mu$ = 5$\times$10$^{-8}$ and $y$ = 10$^{-8}$ at the 5$\sigma$ level therefore requires a highly refined foreground subtraction, which is further complicated by the fact that the CO signal is not completely spectrally smooth. At higher frequencies, $\nu \gtrsim$ 400 GHz, the foreground CO emission grows exponentially weak, $\Delta I_\nu^{\rm{CO}} \ll$ $\Delta I_\nu^{\mu}$, $\Delta I_\nu^{y}$, and ceases to pose as a prominent limiting factor in obtaining accurate spectral distortion measurements. 
The $y$ distortion from reionization and structure formation ($z \lesssim$ 10 - 20; green curve), has a relatively large amplitude, $|y| \sim$ 2$\times$10$^{-6}$, comparable to the predicted cumulative CO signal, and is expected to be visible even without more detailed modeling.

\section{Discussion}
We apply an LVG-based modeling approach to predict the cumulative CO emission signal generated by star-forming galaxies throughout cosmic history. The relative CMB distortion due to this CO foreground is not spectrally smooth, but rather peaks in the frequency range $\nu \sim$ 100 - 200 GHz with an amplitude of $\Delta I_\nu/I_\nu \simeq$ 5$\times$10$^{-6}$. Exploring cases where nearby sources of CO emission can be identified and subtracted from observations, we find that the dominant contributors to the CO signal originate from star-forming halos with masses $M \sim$ 10$^{11}$-10$^{12}$ M$_\odot$ at modest redshifts of $z \sim$ 2 - 5. 
While the intensity of this cumulative CO foreground is at least 2-100 times weaker than the current COBE/FIRAS upper limits, and has thus far evaded detection, it falls well above the detection limit of future instruments, such as PIXIE, which promise to measure CMB spectral distortions with sensitivity improved by 2-3 orders of magnitude compared to COBE/FIRAS. 

In standard cosmology, there are a number of different heating/cooling processes in the early Universe which may have given rise to CMB spectral distortions of varying magnitudes and shapes. 
Silk damping of small-scale perturbations in the primordial baryon-electron-photon fluid is one of them, resulting in CMB distortions with magnitudes of $\Delta I_\nu/I_\nu \simeq$ 10$^{-8}$ - 10$^{-10}$, depending on the shape and amplitude of the primordial power spectrum at scales 50 $\leq k \leq$ 10$^4$ Mpc$^{-1}$\citep{1991ApJ...371...14D,1994ApJ...430L...5H,2012MNRAS.425.1129C}. 
Residual annihilation of dark matter particles throughout the history of the Universe is another, releasing energy that leads to $\mu$ and $y$ distortions of amplitude $\mu \approx$ 3$\times$10$^{-9}$ ($z >$ 5$\times$10$^4$) and $y \approx$ 5$\times$10$^{-10}$ ($z <$ 5$\times$10$^4$), respectively \citep{2012MNRAS.419.1294C,2014MNRAS.438.2065C}. The cosmological recombination of hydrogen and helium is expected to have introduced distortions as well, with amplitudes of $\Delta I_\nu/I_\nu \simeq$ 10$^{-9}$ at redshifts $z \sim$ 1100 - 6000 \citep{2006A&A...458L..29C,2006MNRAS.371.1939R,2008A&A...485..377R}. 
Similar magnitude but opposite sign distortions, $\mu \sim$ -2.7$\times$10$^{-9}$ and $y \sim$ -6$\times$10$^{-10}$, are expected from energy losses of the CMB to baryons and electrons as they cool adiabiatically faster than radiation with the expansion of the Universe \citep{ChlubaPhD,2012MNRAS.419.1294C,2012A&A...540A.124K}.

Experiments like PIXIE will be able to constrain these spectral distortions in the CMB at the 5$\sigma$ level, providing measurements at sensitivities $\mu$ = 5$\times$10$^{-8}$ and $y$ = 10$^{-8}$. However, as demonstrated above, the cumulative CO foreground is an important contaminant to these cosmological distortions, with a signal strength, $\Delta I_\nu^{\rm{CO}}/I_\nu \sim$ 5$\times$10$^{-6}$- 10$^{-7}$, that is 1-3 orders of magnitude higher than PIXIE's sensitivity limits in the frequency range $\nu \sim$ 20 - 360 GHz.
Based on the results depicted in Figure 2, CO luminous sources at redshifts $z <$ 8 need to be identified and subtracted in order to reduce this cumulative signal to levels that are at least comparable to the $\mu$- and $y$-type spectral distortions one hopes to constrain. Removing all such sources is challenging, both in terms of exposure time and in terms of field coverage. Even with ten hours of integration time, instruments like the Atacama Large Millimeter Array (ALMA) will miss CO emission from host halos with masses $M \lesssim$ 5$\times$10$^{10}$ M$_\odot$, which contribute tens of percent of the cumulative CO signal at redshifts $z \gtrsim$ 4. (In this paper, we integrate over the population of CO luminous halos with masses $M \geq$ 10$^{10}$ M$_\odot$ and thus present conservative estimates of the total CO foreground which do not account for contributions from lower mass halos, $M \lesssim$ 10$^{10}$ M$_\odot$.) 
Removing the aggregate line emission from unresolved sources throughout cosmic history poses its own difficulties.
Unlike the spectrally smooth synchrotron and thermal dust foregrounds which can be approximately described by power laws, the CO foreground fluctuates in frequency due to the clustering of sources over restricted regions on the sky; accurate foreground subtraction therefore requires knowledge of the emission spectrum to high order of precision, challenging our ability to fully exploit PIXIE's sensitivity to constrain CMB spectral measurements.

\section{Acknowledgements}
This research was also supported by the Raymond and Beverly Sackler Tel Aviv University - Harvard/ITC Astronomy Program. This work was supported in part by NSF grant AST-1312034. This material is based upon work supported by the National Science Foundation Graduate Research Fellowship under Grant No. DGE1144152. Any opinion, findings, and conclusions or recommendations expressed in this material are those of the authors and do not necessarily reflect the views of the National Science Foundation.

\label{lastpage}
\end{document}